# Rapid CVD growth of millimetre-sized single crystal graphene using a cold-wall reactor

Vaidotas Miseikis[1], Domenica Convertino[1,2], Neeraj Mishra[1], Mauro Gemmi[1], Torge Mashoff[1], Stefan Heun[3], Niloofar Haghighian[4], Francesco Bisio[5], Maurizio Canepa[4], Vincenzo Piazza[1] and Camilla Coletti[1,2]

[1]Center for Nanotechnology Innovation @NEST, Istituto Italiano di Tecnologia, Piazza San Silvestro 12, 56127 Pisa, Italy
[2]Graphene Labs, Istituto Italiano di Tecnologia, Via Morego 30, 16163 Genova, Italy
[3]NEST, Istituto Nanoscienze - CNR and Scuola Normale Superiore, Piazza San Silvestro 12, 56127 Pisa, Italy
[4]Dipartimento di Fisica, UniGe, Via Dodecaneso 33, 16146 Genova, Italy
[5]CNR-SPIN, Corso Perrone 24, 16152 Genova, Italy

E-mail: vaidotas.miseikis@iit.it, camilla.coletti@iit.it

**Abstract**
In this work we present a simple pathway to obtain large single-crystal graphene on copper (Cu) foils with high growth rates using a commercially available cold-wall chemical vapour deposition (CVD) reactor. We show that graphene nucleation density is drastically reduced and crystal growth is accelerated when: i) using ex-situ oxidised foils; ii) performing annealing in an inert atmosphere prior to growth; iii) enclosing the foils to lower the precursor impingement flux during growth. Growth rates as high as 14.7 and 17.5 $\mu$m per minute are obtained on flat and folded foils, respectively. Thus, single-crystal grains with lateral size of about one millimetre can be obtained in just one hour. The samples are characterised by optical microscopy, scanning electron microscopy (SEM), X-ray photoelectron spectroscopy (XPS), Raman spectroscopy as well as selected area electron diffraction (SAED) and low-energy electron diffraction (LEED), which confirm the high quality and homogeneity of the films. The development of a process for the quick production of large grain graphene in a commonly used commercial CVD reactor is a significant step towards an increased accessibility to millimetre-sized graphene crystals.

## 1. Introduction

Graphene produced by micromechanical exfoliation is still preferred for most fundamental research due to perfect crystallinity and low density of defects. However, the typical small size makes the exfoliated flakes unsuitable for large-scale graphene-device production of any kind. The two main growth techniques of high-quality wafer-scale graphene are the thermal decomposition of silicon carbide (SiC), whereby graphene is produced directly on the surface of insulating SiC wafers[1,2] or chemical vapour deposition (CVD) growth of graphene on transition metals. Nickel is a common substrate for the growth of multi-layer films of graphene[3]. Due to appreciable solubility of carbon in nickel (0.6% by weight at temperatures above 1000 °C)[4], carbon diffuses into the metal and then segregates to form multi-layer films of graphene. The segregation process is difficult to control, therefore such films are generally not very homogeneous and large areas of single- or bi-layer graphene are hard to achieve. On the contrary, the solubility of carbon in copper (Cu) is very low (2 orders of magnitude lower than that in nickel)[4], and therefore the growth occurs largely by a surface mediated process[5]. This allows for the production of homogeneous single-layer films, making copper by far the most common substrate for CVD growth of graphene. During the growth process, hydrocarbons adsorbed on the copper surface are dissociated and become highly mobile, forming ordered crystals of graphene[6]. In the presence of oxygen, the growth of graphene crystals can be accelerated by lowering the edge attachment energy for carbon atoms[7]. While the total size of the graphene samples grown on copper are only limited by the size of the substrate[8], the CVD-grown films are polycrystalline, with a typical grain size of around 10 $\mu$m, depending upon the growth conditions. It has been shown that the properties of polycrystalline films of CVD graphene can be significantly inferior to those of single crystals[9]. While recent theoretical studies suggest that the grain boundaries themselves may not have a strong intrinsic effect on carrier transport, they are the sites of preferable adsorption of external species, which can cause appreciable scattering of charge carriers[10]. One of the primary goals in the recent studies of CVD growth of graphene has been to increase the size of single crystals. By limiting the flow of the carbon precursor[11,12] or passivating the Cu surface in-situ[7], single grains of graphene as large as several millimetres in diameter have been prepared[7,12–15], however, such samples are typically grown in horizontal quartz tube systems and often require long growth processes[12–15]. In addition, the great variability in the configuration of home-made CVD reactors makes it hard to transfer process parameters from one system to the other. Moreover, large grain graphene is often obtained adopting the so-called Cu "pocket" configuration, originally shown in reference 11. Their formation requires manual folding of the foils and their implementation can be cumbersome and hard to reproduce. Also, the subsequent foil unfolding can introduce mechanical stress on the grown graphene.

In this work we present a growth strategy which allows for the quick production of large crystals of graphene not only on Cu "pockets", but also directly on flat foils. The growth process is implemented in a commercially available reactor and easy to reproduce. We demonstrate that by annealing oxidised Cu foils in an inert atmosphere and by employing a simple sample enclosure to limit the impingement flow, we can reduce the nucleation density by about four orders of magnitude. Single crystals with a grain size approaching one millimetre can be grown in just one hour. Using the copper "pocket" configuration, grains with lateral dimensions up to 3.5 mm are obtained in 3 hours. The relatively fast ramp-up and cool-down rates of our cold-wall reactor, along with a short annealing process, add only 1.5 hours to the growth time thus contributing to a significantly reduced total process time (e.g. 2.5 hours for the production of 1 mm single-crystals).

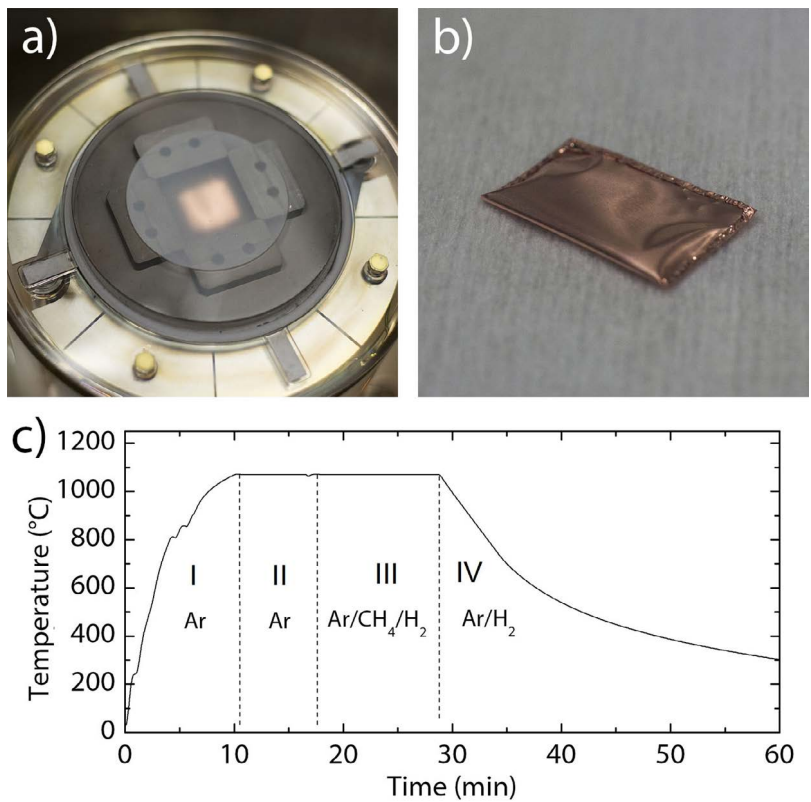

**Figure 1.** a) Quartz/graphite sample enclosure; b) Copper "pocket" enclosure; c) Temperature profile of a typical 4-stage CVD growth process. (I) Temperature ramp-up, (II) Annealing, (III) Growth, (IV) Cool-down.

## 2. Methods

### 2.1. Cu surface preparation and CVD growth

The 25-$\mu$m thick Cu foil used in this work (unless specified otherwise) was supplied by Alfa Aesar (purity 99.8%, lot no. 13382). Such foil is also referred to as low-purity in the text. The foil was electropolished in an electrochemical cell made using a commercially available Coplin staining jar as the vessel and an electrolyte solution described previously[12]. The grooves of the staining jar ensured that the foil was kept flat and parallel to the counter electrode (a thicker Cu plate), which helped achieve homogeneous polishing of the surface. More detailed information on the electropolishing setup and process as well as its effects can be found in the supplementary information (SI).

Graphene films were synthesised at a pressure of 25 mbar inside a 4-inch cold-wall CVD system (Aixtron BM). To achieve more reproducible atmospheric and thermal conditions and to reduce the effective gas flow, the sample was contained in a custom-made enclosure, comprising a quartz disk suspended 6 mm above the sample using graphite spacers (Fig. 1 (a)). A typical temperature profile of a CVD growth process is shown in panel (c) of Figure 1, indicating the 4 distinct parts: temperature ramp-up (I), annealing (II), growth (III) and cool-down (IV). The annealing as well as the growth was performed at approximately 1060 °C, which was calibrated according to the melting point of Cu. The annealing time was kept at 10 minutes in all cases. The gas flow during the temperature ramp-up and the annealing stages (panel (c), I and II) was 1000 sccm. The samples which had the largest grains were annealed in argon atmosphere (mentioned as Ar-annealing process in the text); other samples were annealed in hydrogen (mentioned as hydrogen annealing in the text). The gas flow rates during growth (panel (c), III) were typically set to 1 sccm of methane, 20 sccm of hydrogen and 900 sccm of argon. The growth time varied depending on the nucleation density, from 3 minutes to 3 hours. After the growth, the chamber was cooled in argon/hydrogen atmosphere to a temperature of 120 °C before introducing the samples to air (Fig 1. (c), IV).

### 2.2. Characterisation

After each growth process, assessment of the graphene coverage and analysis of the size and the shape of the graphene grains was performed using optical and scanning electron microscopy (SEM). The largest isolated grains grown after Ar-annealing were clearly visible on the Cu substrates using optical differential interference contrast (DIC) imaging carried out with a Nikon AZ100 microscope. In the case of samples grown adopting hydrogen annealing, the contrast was generally lower and the graphene was only seen using SEM (Zeiss Merlin, operating at 5 kV). Raman spectroscopy was used to analyse the quality of the synthesised graphene after transferring it on Si/SiO$_2$ substrates, as described in the SI. Raman analysis was performed using a Renishaw InVia system equipped with a 532 nm green laser and a motorised stage for large-area mapping. A laser spot size of approximately 1 $\mu$m in diameter was used. Transmission electron microscopy (TEM) analysis was carried out on a Zeiss Libra 120 TEM operating at 120 kV and equipped with an in-column Omega filter. The crystallinity of graphene and its relation with the underlying copper substrate was analysed using LEED measurements, which were performed in an ultra-high vacuum environment with a base pressure below 1 x 10$^{-10}$ mbar using 6-inch LEED/Auger optics (OCI Vacuum Microengineering). Chemical characterization of the foil before and after growth was performed using XPS (PHI ESCA 5600 system, monochromatised Al source). The energy scale of the core level spectra was calibrated by reference to the adventitious C1s peak at 284.5 eV measured before graphene growth.

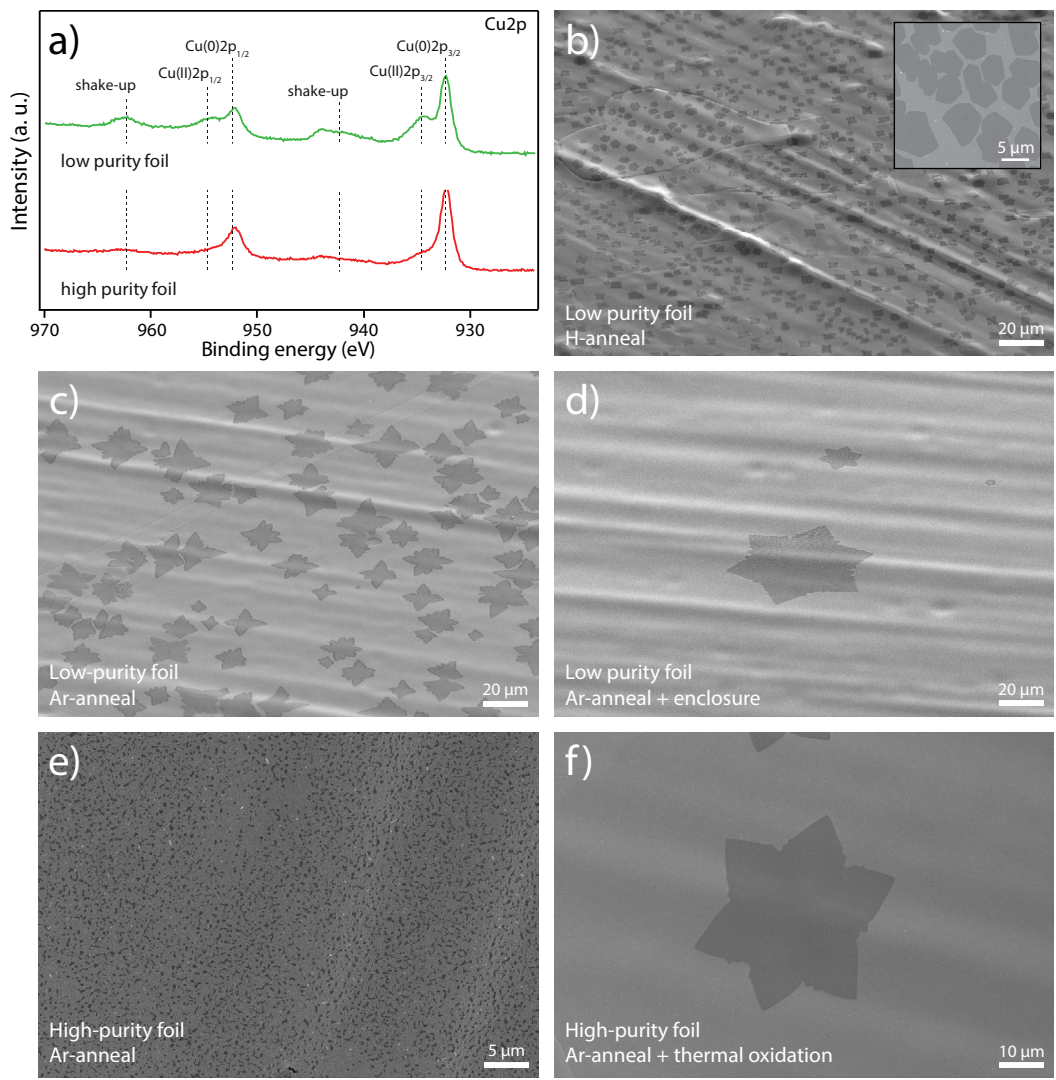

**Figure 2.** (a) XPS spectra of the Cu2p core-level emission region taken before graphene growth on low-purity and high-purity Cu foil. (b) A tilted SEM image of graphene grown on low-purity Cu foil annealed in hydrogen. Inset: top-view image showing the compact edge grains. (c) Tilted image of graphene grown on low-purity Cu foil annealed in argon without sample enclosure. (d) Tilted image of graphene grown on low-purity Cu foil annealed in argon with sample enclosure. (e) Small-crystal film grown on high-purity copper foil with low oxygen content. (f) SEM image of a large isolated single crystal grown on high-purity Cu foil after intentional oxidation at 180 °C.

## 3. Results and discussion

### 3.1. Combining surface-passivated foil and Ar annealing: a path towards millimetre size grains on flat foil

Chemistry of the Cu surface has been recently shown to play an important role, strongly influencing nucleation density and growth rate[7]. In this work, we perform chemical characterization of the Cu foil via XPS. The top spectrum in Fig. 2 (a) shows the Cu 2p core-level emission region of the Cu foil measured after electropolishing and before growth. A line-shape analysis reveals several components. The most prominent Cu doublets - Cu $2p^{3/2}$ (932.45 ±0.15 eV) and Cu $2p^{1/2}$ (952.4 ±0.15 eV) - are assigned to bulk Cu. Intense side peaks are located at 934.7 ±0.2 eV and 954.2 ±0.2 eV. According to literature[16], these shoulders are assigned to the $Cu^{+2}$ state (CuO). As expected, prominent shake-up satellites of $Cu^{+2}$ are found at about 9 eV above the main $Cu^{+2}$ doublet[17]. This analysis evidences the presence of a native cupric oxide on the Cu foil surface.

During preliminary experiments, graphene was grown on such foils using a classical growth process with hydrogen annealing in a configuration without sample enclosure. Complete graphene coverage of the surface was observed after just 5 minutes of growth (characterization via SAED and TEM mapping in the SI). Partial coverage experiments with a growth time of 1 minute revealed a high nucleation density (≈ 12500 grains per mm$^2$) with an average grain size of a few microns (Fig. 2(b)). The nucleation of graphene on hydrogen-annealed samples occurred primarily on the surface irregularities such as the rolling grooves which are typically formed during the production of Cu foils. Indeed, hydrogen annealing of uneven Cu surfaces formed crystal terraces with sizeable steps, which acted as energetically favourable spots for the nucleation of graphene[18]. During annealing in hydrogen, the cupric oxide is reduced to metallic copper[19]: thus, the nucleation is dense and concentrated on the energetically favourable surface features. We also note that hydrogen annealing produced Cu single crystal domains with a limited lateral size (i.e., about 50-100 $\mu$m).

By simply replacing hydrogen annealing with Ar-annealing, the nucleation density was reduced by one order of magnitude (i.e., from about 12500 to about 1000 grains per mm$^2$ after 1 minute growth). Larger grains - with sizes up to 20 $\mu$m - could be obtained (see panel (c)). Also, preferential nucleation along the direction of the Cu grooves was not observed. While for samples obtained with hydrogen annealing the grain growth fronts were compact (inset in Fig. 2 (b)), dendritic fronts were observed when annealing in argon (Fig. 2 (c) and (d)). Indeed, dendritic edges are typical for diffusion-limited growth, the dominant growth mechanism when oxygen is preadsorbed on the copper surface and mediates the growth process[7]. Clearly annealing in an inert atmosphere prevents reduction of the native oxide. Thus, similarly to what was reported in ref. 7 for in-situ oxidised copper, by maintaining the native oxide up to growth initiation, carbon nucleation is suppressed and crystal growth accelerated. Notably, Ar-annealing produced much larger Cu single crystal domains, which were on the order of several millimetres.

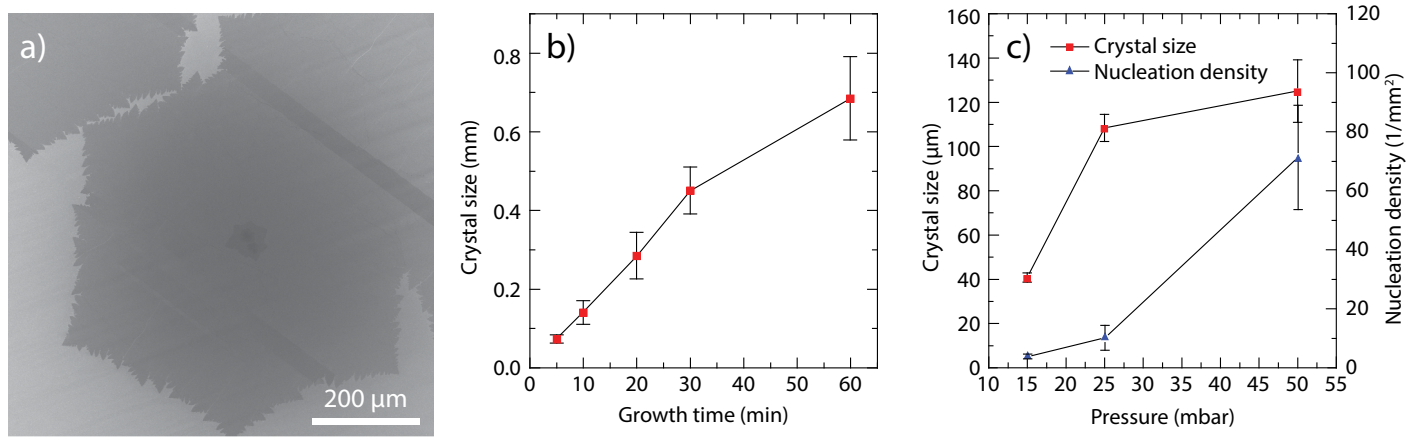

**Figure 3.** (a) SEM image of an isolated graphene crystal with a diagonal dimension of nearly 750 µm. (b) The size of graphene crystals as a function of growth time. (c) Crystal size and nucleation density obtained after 10 minutes of growth as a function of process pressure.

When non-reducing Ar-annealing was used in combination with the simple sample enclosure shown in Fig. 1 (a), further progress in reducing the nucleation density could be observed. The nucleation of graphene after 1 minute of growth was nearly three orders of magnitude sparser than that measured for the non-enclosed configuration. Figure 2 (d) shows a typical crystal formed after 5 minutes of growth inside a sample enclosure, with a lateral size of about 65 µm. Indeed, confining the sample in a reduced volume limits the net number of carbon precursor species accessible to the Cu foil, with the consequence of reducing the nucleation density. It should be mentioned that all process gases used in this work had high purity (99.9999%). Thus, it is reasonable to assume that the surface reaction was dominated by the oxygen initially contained in the copper foil. To rule out the possibility that oxygen residue in the reactor (estimated value is calculated and presented in the SI) is responsible for such diffusion-limited growth, we performed two sets of experiments. Firstly, we annealed the Cu foil at 1000 °C and $10^{-3}$ mbar in the reactor just before initiating growth. Such thermal treatment should lead to desorption of the native oxide and of the oxygen species segregated from the bulk[19]. Vacuum annealing was followed by the standard Ar-annealing process adopted for the samples shown in Fig. 2 (c). This yielded high nucleation density (comparable to the samples produced using hydrogen annealing, such as the one shown in panel (b)). As an additional test, growth experiments were also performed using high-purity Cu foil (Sigma-Aldrich, 99.98%). XPS analysis of such foil performed after electropolishing and before growth (bottom spectrum in Fig. 2 (a)) revealed a much lower CuO shoulder than that observed for the low-purity foils (top spectrum in Fig. 2 (a)), thus indicating a lower level of cupric oxide at the surface. Using the standard Ar-annealing process the foils were covered with a continuous carbon film in as low as 1 minute of growth. By reducing the growth time to 45 seconds, non-continuous films were grown with a disordered structure (Fig. 2 (e)), further confirming the nanocrystalline nature of this material. Samples of high-purity foil were then intentionally oxidised by exposing them to ambient atmosphere at 180 °C for 2 minutes. Growth performed on the oxidised high-purity foil was comparable to the low-purity foil with a relatively low nucleation density. Five minutes of graphene growth produced isolated crystals with a diameter of ~60 µm (figure 2 (f)). These results (together with the calculations presented in the SI) indicate that it is the oxygen initially contained in the foil, rather than the oxygen residue in the reactor, which plays a relevant role in the formation of large grain graphene. Clearly, using a non-reducing annealing is vital to preserve oxidation up to growth initiation, thus positively affecting the nucleation density.

### 3.2. Large grain graphene on flat foil

As suggested in the previous section, oxygen passivated foil and Ar-annealing in an enclosed environment is the combination which yields the lowest nucleation density and conclusively larger single-crystal graphene grains in short times. To determine the optimum growth conditions, nucleation density determined over large areas as well as crystal size for a given growth duration was analysed as a function of process pressure. The results of these experiments are shown in Figure 3 (c). The optimum process pressure was found to be 25 mbar, which allowed obtaining a relatively high growth rate while maintaining a reasonably low nucleation density. 50 mbar process yielded a nucleation density as high as 70 grains per mm2, ruling out the growth of large grains. Despite the low nucleation density obtained at 15 mbar, this process was not found to be suitable for the rapid growth of large crystals due to the low growth rate. Additionally, the ideal geometrical configuration of the sample enclosure was investigated and an enclosure height of 6 mm was found to yield the best results (detailed studies are presented in the SI). Using these growth conditions, single grains with a diagonal size up to 800 µm could be grown in 1 hour (Fig. 3 (a) and (b)). The crystals had dendritic edges and showed a clear six-fold symmetry. For growth durations ranging from 5 to 30 minutes the dependence of crystal size versus growth time is linear, with an estimated growth rate of ≈ 14.7 µm per minute. An average lower growth rate (i.e., 11.4 µm per minute) and larger error bars are observed for growth times of 60 minutes: this is due to the merging of neighbouring crystals observed at this stage.

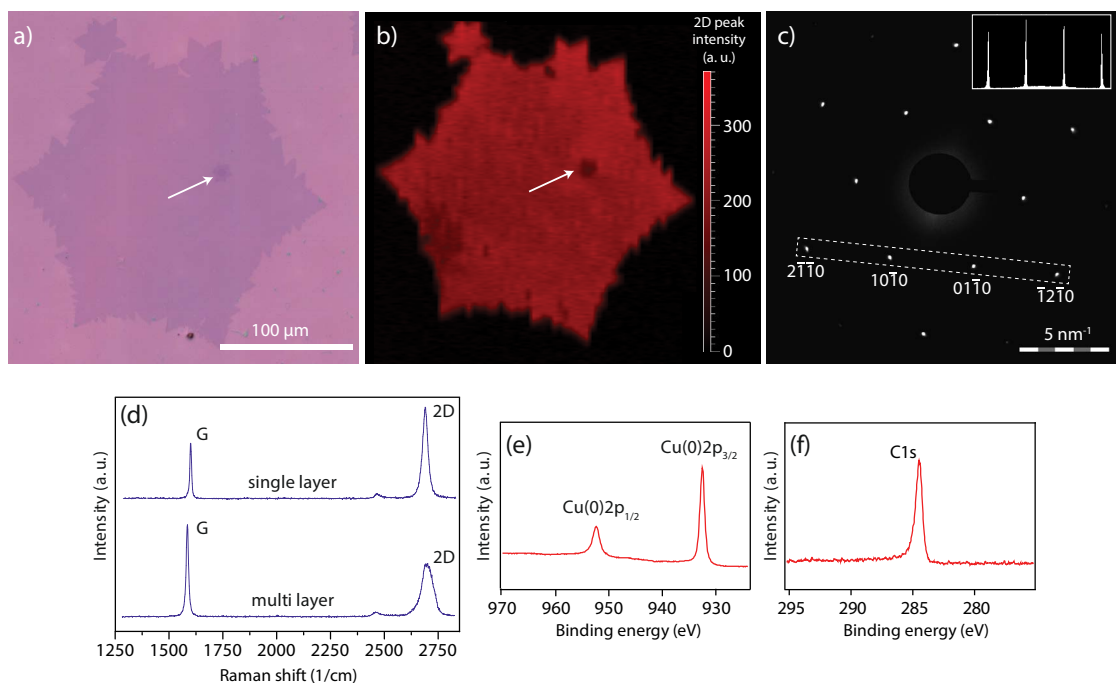

**Figure 4.** a) Optical image and b) intensity map of Raman 2D peak of a large single grain transferred to Si/SiO2. c) Typical SAED pattern of large grain graphene with the inset showing the intensity plot of the SAED spots indicated by the dashed rectangle. (d) Raman spectra taken from a single-layer (top) and multi-layer (bottom) region of an isolated grain. XPS spectra measured after graphene growth for (e) Cu2p and (f) C1s core-level emission regions.

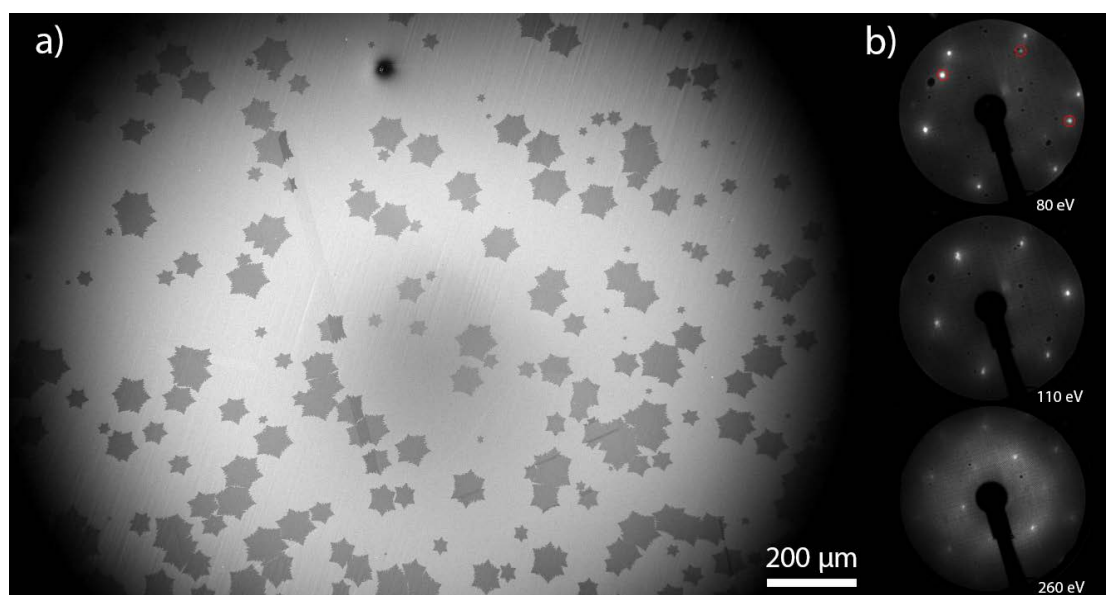

**Figure 5.** a) Low-magnification SEM image at the early stage of large-grain graphene growth on flat foil. The nucleation density is around 20 grains/mm2. b) LEED patterns of graphene on Cu taken at different electron energies. At 80 eV, an additional set of diffraction spots appears due to the faceted copper surface. 3 of the spots are within the field of view and are indicated by red circles.

### 3.3. Structural and chemical characterization of large grain graphene

Raman spectroscopy was performed on large single crystals such as the one shown in Fig. 4 (a). A typical Raman spectrum acquired for this sample is reported in the top of Figure 4 (d) and is characteristic for single-layer graphene with prominent G and 2D peaks[20]. The absence of the D-peak and a narrow 2D peak (FWHM = 29 cm$^{-1}$) confirm the high crystal quality of the syntesised graphene[21]. The bottom Raman spectrum in Figure 4 (d) is typical for the small multi-layer graphene patches, which were sometimes observed on the single grains (white arrow in panels (a) and (b)). As expected for Bernal-stacked graphene, the 2D peak is much broader (FWHM = 62 cm$^{-1}$) and the G/2D intensity ratio is larger than for single-layer graphene[20]. To confirm the high quality of graphene and the single-layer nature of the grains over large areas, high-resolution Raman mapping was performed. The measured 2D peak intensity map is shown in Figure 4 (b). The intensity is very homogeneous across the whole grain except for the small multi-layer patch. The intensity of the D-peak was negligible across the whole sample area mapped.

Selective-area electron diffraction (SAED) analysis of the graphene transferred to TEM grids was performed at multiple points of the samples. The SAED patterns obtained on large-grain graphene using the largest SAED diaphragm corresponding to an illuminated area of 4.5 μm in diameter were clear and sharp, as shown in Figure 4 (c), and no significant rotation of the patterns was observed over distances of hundreds of microns, confirming the large size of single crystals. While small (1-2°) rotation of the diffraction spots was sometimes observed within the single-grains, it was most likely caused by the wrinkles in graphene, which are typically formed during transfer as well as the post-growth cooling[5]. The inset of Figure 4 (c) shows the intensity profile of inner (01-10), (10-10) and outer (2-1-10), (-12-10) SAED spots. The relative spot intensities are consistent with the theoretical and experimental results previously obtained for suspended single-layer graphene[22].

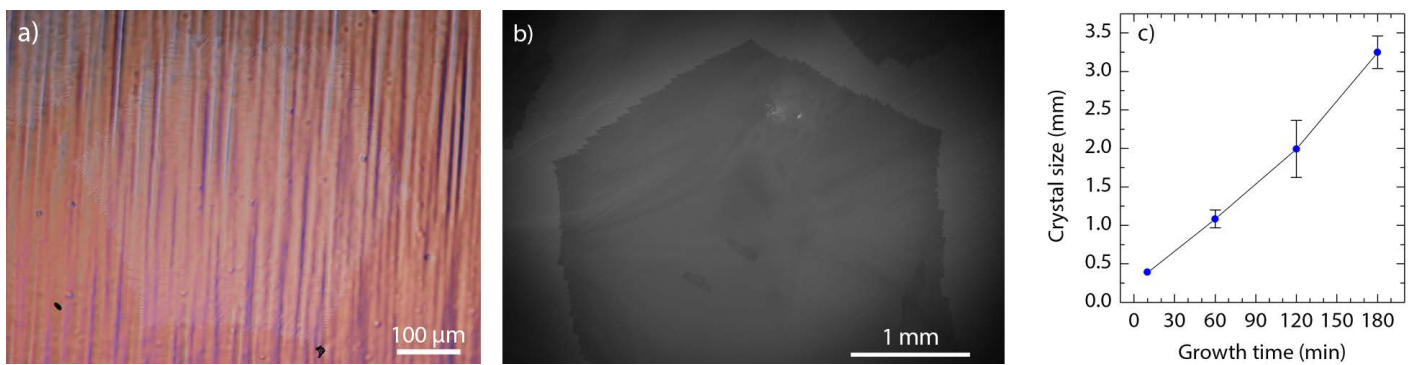

**Figure 6.** Large single grains produced inside a copper "pocket". a) Optical DIC image of a hexagonal grain on copper. A sub-millimetre grain was chosen to fit into the field-of-view of the microscope. b) SEM image of an single grain with a diameter of over 3 mm. c) the size of grains as a function of growth time.

It should be pointed out that even for short growth times and higher nucleation densities than those observed in optimum conditions, the separate grains grown on the same Cu crystal domain typically presented an aligned crystal orientation, which suggests epitaxial growth on the surface of copper (see Fig. 5 (a)). Rotational alignment of graphene grown on copper has been studied previously, showing a strong dependence on the crystal structure of the copper substrate. In the case of copper (111), the honeycomb graphene lattice is expected to align perfectly with the copper crystal, with no rotational mismatch[23,24]. For other crystal structures of copper, such as Cu (100), which is commonly observed in annealed copper foils[24,25], graphene is expected to only have weak interaction with the substrate due to the square lattice of the copper. Nonetheless, a preferential growth orientation can be still observed due to the alignment of graphene lattice to copper facets[25]. Further analysis of the crystal orientation of our samples was performed using LEED, employing an electron beam of approximately 0.5 mm in diameter. The obtained diffraction patterns showed a predominantly hexagonal symmetry, indicating formation of mainly Cu (111) crystals during the annealing. The hexagonal patterns maintained rotational alignment at all measured energies, ranging from 80 eV to 260 eV (Fig. 5 (b)), which indicates epitaxial growth with no rotational mismatch. At 80 eV the LEED signal is dominated by graphene, whereas at 260 eV the substrate plays a major role and a 3-fold symmetry is revealed[24]. In some cases, additional diffraction spots were observed, forming shifted hexagonal patterns due to the faceted copper surface, as previously observed by Hao *et al*[7]. Three of such spots are visible in the top panel of Figure 5 (b) and are indicated by red circles (the other three spots are outside of the field of view). The positions of these secondary spots were found to shift with electron energy and they matched the main spots at 110 eV (Figure 5 (b), middle).

Finally, XPS analysis was performed after graphene growth. In the C1s core level spectrum in Fig. 4 (f) only the graphene-related peak is visible and no additional component indicative of adventitious carbon is observed. In agreement with literature, the graphene peak could be well fitted with a Doniach-Sunjic line shape, accounting for the metallic behaviour (asymmetry value 0.12)[26]. The peak position at 284.5 eV indicates that the graphene grown is neutrally charged[26]. The measured Cu 2p spectrum presented only the two sharp peaks characteristic of metallic Cu (Cu$^0$) (Fig. 4 (e)). Shoulders belonging to cupric oxide were not observed. This result is consistent with the model presented in reference 7: oxygen species are consumed and desorbed during the attachment and incorporation of carbon atoms.

### 3.4. Quick production of large grain graphene in Cu "pockets"

In this work we also investigate the growth of graphene inside the Cu "pocket" configuration shown in Fig. 1 (b) and previously described by Li *et al*[11]. The copper surface on the inside of such "pockets" has highly limited exposure to gaseous carbon species and therefore the nucleation density was reduced to well below 1 grain per mm$^2$ in our case. This allowed us to grow isolated crystals with a size of 1.2 mm in 1 hour and up to 3.5 mm in 3 hours (Figure 6). Figure 6 (c) shows the size dependence versus growth time for graphene crystals grown inside the copper "pockets". Using the same basic parameters, the growth rates are slightly higher than those observed for flat foils, at an average of ≈ 17.5 $\mu m$ per minute.

We note that copper "pockets" allowed us to achieve millimetre-sized grains without the use of the quartz/graphite enclosure; however the quartz "roof" provided a more practical benefit of drastically reducing the contamination of the CVD chamber by the Cu sublimation at high temperatures, a common issue for the low-pressure CVD growth on copper[12]. The cleaning procedure for the disk was much simpler compared to the whole CVD chamber and thorough cleaning could be performed after each process. As we found experimentally, cleanliness of the chamber is a crucial condition for obtaining samples with low nucleation density. Although larger crystals can be obtained by using "pockets", it should be considered that the folding and opening of the foil is not necessarily suitable for the routine preparation of samples for applications. Considering the typical dimensions of graphene-based devices, the crystal dimensions achieved on flat foil should satisfactorily comply.

## 4. Conclusion

In this work we present a novel combined strategy which allows for the production of large crystals of graphene with high growth rates. We show that by using ex-situ passivated Cu foil (i.e., either with a native or thermally grown cupric oxide) and by maintaining the oxide up to growth initiation in a non-reducing environment, nucleation density is significantly reduced and crystal growth accelerated. We demonstrate that by employing a simple sample enclosure, crystals approaching 1 mm in lateral size can be obtained in just 1 hour directly on flat foils. When adopting a "pocket" configuration 3.5 mm grains can be obtained in 3 hours. Optical and electron microscopy, Raman spectroscopy, and selected-area electron diffraction confirm the high crystallinity and homogeneity of the synthesised films. Low-energy electron diffraction measurements reveal that the rotational alignment is also maintained between separate crystals of graphene, indicating epitaxial growth on the copper surface. XPS analysis evidences a complete removal of the initial cupric oxide after graphene growth and indicates that the grown layer is neutrally charged. The detailed report of a method which allows for the quick production of high-quality large-area graphene in a commonly used commercial CVD reactor is likely to lead towards an increased accessibility to millimetre-sized graphene crystals and, ultimately, a wider use of CVD graphene in electronic and photonic applications.

## Acknowledgments

The authors would like to thank: V. Voliani and G. Signore of CNI@NEST, IIT, Pisa for assistance with electropolishing of the Cu foil and for useful discussions; K. Teo and N. Rupesinghe from Aixtron for technical support with the Aixtron BM system: G. Gemme of INFN, Sezione di Genova for access to the XPS facility. F. B., M. C. and N. H. acknowledge financial support from the Ministero dell'Università, dell'Istruzione e della ricerca (PRIN 20105ZZTSE_003). The research leading to these results has received funding from the European Union Seventh Framework Programme under grant agreement n°604391 Graphene Flagship.

# Supplementary information

## 1. Cu foil preparation: electropolishing

To prepare the copper foil for the growth of graphene, the foil was electropolished in a home-made electrochemical cell, which used an ISO-TECH bench top power supply as the DC voltage source and a Coplin staining jar as the vessel. The set-up is shown in Figure S1. The copper foil was used as the anode and a thicker copper plate was used as the cathode, which were connected to the power supply using crocodile clips. The electrolyte solution was a mixture of 25 mL of water, 12.5 mL of phosphoric acid, 12.5 mL of ethanol, 2.5 mL of isopropanol and 0.4 g of urea.

A constant voltage of +7 V was applied to the foil in respect to the counter electrode for 60 seconds, producing a current of approximately 1.5-2 A, depending on the sample size. Importantly, the staining jar was chosen for the cell due to the grooves on the inside, which allowed us to keep the foil flat and parallel to the counter electrode (as seen in Fig. S1 (b)), which ensured homogeneous polishing with highly repeatable results.

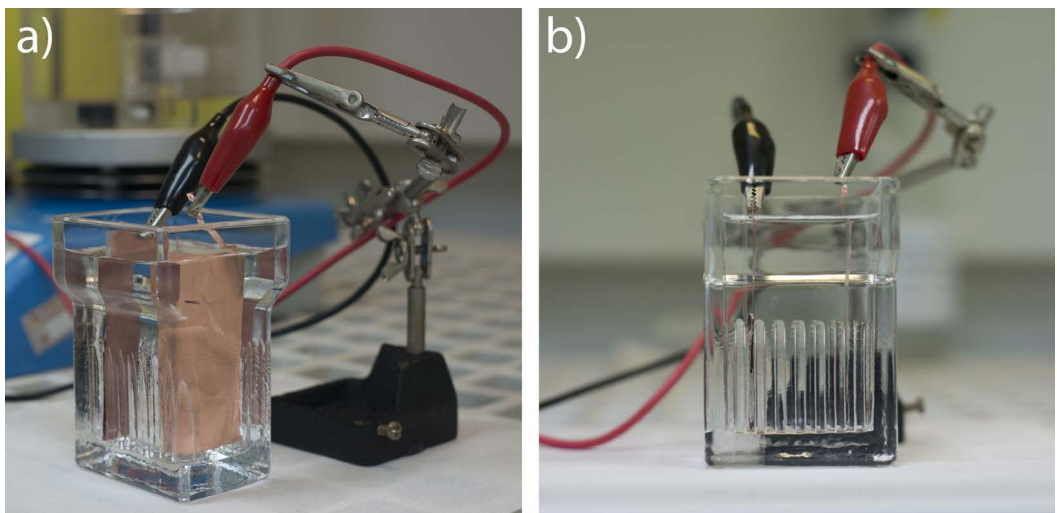

**Figure S1.** a) The electropolishing cell. b) Side profile image shows the parallel sample position ensured by the grooves of the staining jar.

After the polishing, the foil was rinsed with running deionised water and isopropanol. To prevent uncontrollable oxidation of the Cu, the polished foil was kept in isopropanol and dried with compressed nitrogen a few minutes prior to loading into the CVD reactor.

During preliminary experiments, the importance of copper surface treatment was studied. Electrochemical polishing was chosen as a method which not only removes surface contamination, but reduces surface roughness, which can be beneficial for reducing the nucleation density. Growth experiments performed on as-received foil without electropolishing produced samples with a higher graphene nucleation density and large amounts of contaminant particles (Figure S2), which were identified by energy-dispersive X-ray (EDX) analysis to be mostly $SiO_2$. Our electropolishing set-up was designed to provide homogeneous treatment of the foil and therefore a short polishing procedure of only 60 seconds was found to be sufficient to remove nearly all surface contamination and ensure consistent growth of defect-free graphene films over many runs, provided the CVD chamber was kept clean.

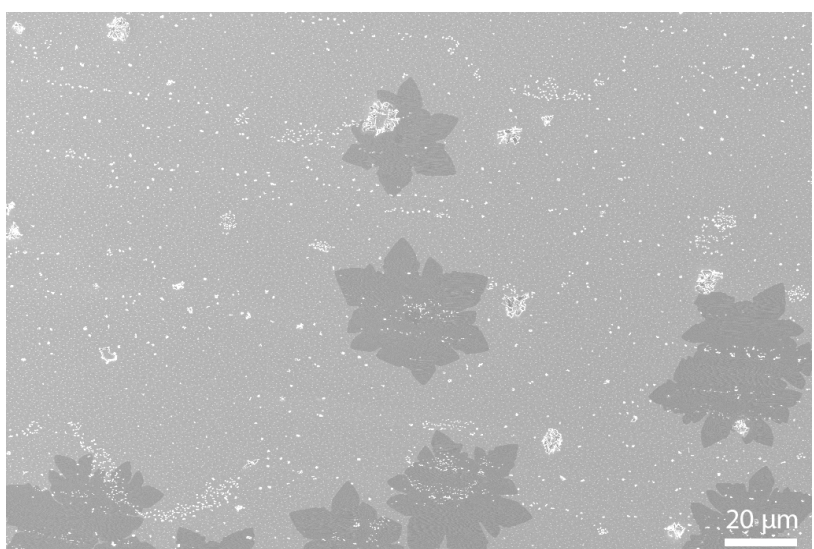

**Figure S2.** CVD growth on untreated copper foil.

## 2. The effect of process pressure: SEM analysis

The growth was performed at several different values of process pressure, analysing the nucleation density and crystal size for a given growth duration in order to determine the optimum growth parameters. The lower limit of 15 mbar was determined by our CVD system (for the given values of the gas flow) and the upper limit of 50 mbar was set by the high nucleation density which made the growth at higher pressures unsuitable for large crystals. As shown in Figure S3, 25 mbar process produced relatively high growth rates while maintaining a low nucleation density, allowing the growth of large single crystals of graphene.

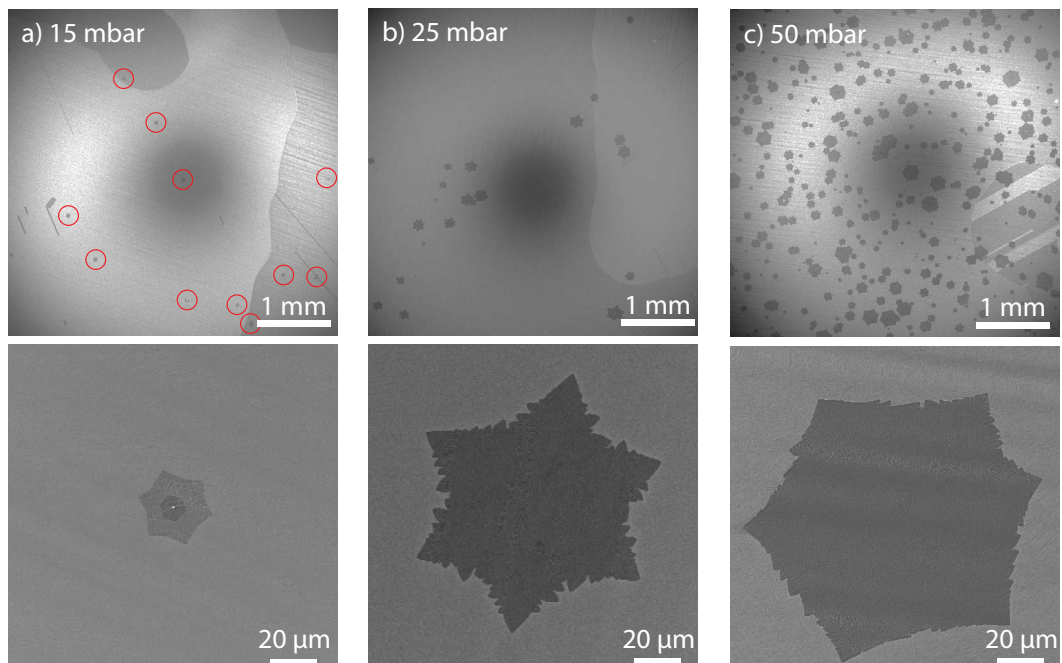

**Figure S3.** SEM images showing the nucleation density and crystal size after 10 minutes of growth under different process pressures. a) Growth performed at 15 mbar. Due to their small size, the nucleating crystals are indicated by red circles. b) Growth performed at 25 mbar. c) Growth performed at 50 mbar.

## 3. Optimisation of sample enclosure geometry

To determine the optimum set-up of the sample enclosure, a series of experiments were performed using different enclosure heights: i.e., we varied the spacing between the sample heater and the quartz disk, which was used as the top of the enclosure (Fig. 1 (a) in the main text). The growth duration was kept at 10 minutes and the graphene nucleation density as well as the size of the largest produced crystals was measured for each different set-up. The results of these experiments are presented in Figures S4 and S5.

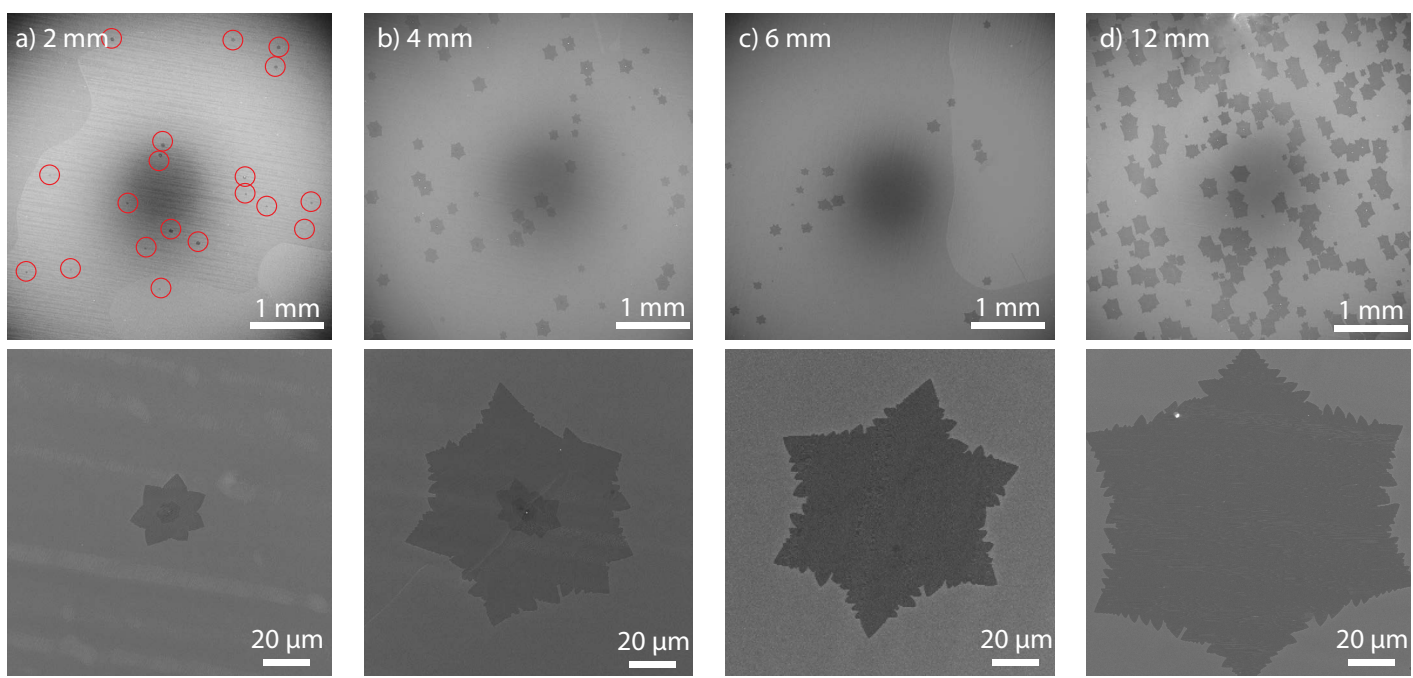

**Figure S4.** SEM images showing the nucleation density and crystal size after 10 minutes of growth using different height of the enclosure. a) Growth performed in a 2 mm enclosure. Nucleating crystals are indicated by red circles. b) Growth performed in a 4 mm enclosure. c) Growth performed in a 6 mm enclosure. d) Growth performed in a 12 mm enclosure.

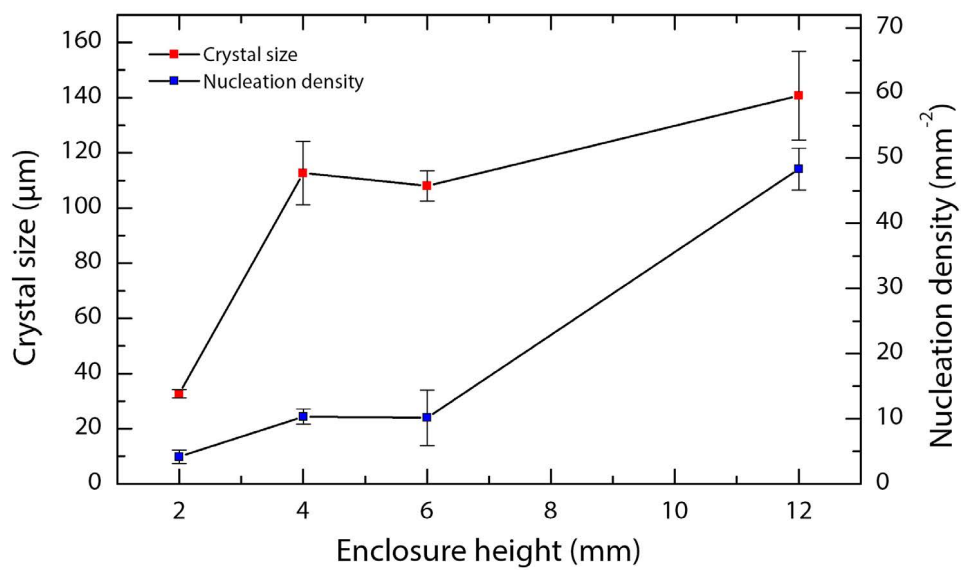

**Figure S5.** Nucleation density and largest crystal size after 10 minutes of growth for different sample enclosure set-ups.

Increasing the height (and consequently the volume) of the enclosure caused a significant increase in nucleation density, which ranged from 5 grains per mm$^2$ using a 2 mm enclosure, to 50 grains per mm$^2$ for a 12 mm enclosure. Very similar results were obtained for the 4 mm and the 6 mm enclosures. The crystal size was also significantly lower using the 2 mm enclosure, but for larger enclosures the results were comparable. In an overview, the 4 mm and the 6 mm enclosure configurations produced similar results and we chose to perform our large grain growth experiments with the 6 mm enclosure due to a simpler set-up.

## 4. Transfer to arbitrary substrates

For Raman spectroscopy and TEM analysis, the CVD-grown graphene was transferred to SiO2/Si substrates and TEM grids (Ted Pella, 2000 and 1500 mesh), respectively. The standard wet poly-methyl-methacrylate (PMMA) transfer techniques were used[S1]. A PMMA solution (4% in anisole) was spin-coated on the foil and left to dry for 1 hour in ambient conditions. The samples were treated with oxygen plasma to remove the back-side graphene coating. The Cu was etched using either a 0.1M solution of iron chloride or 0.44M (100 g/L) solution of ammonium persulfate (Sigma-Aldrich) and the graphene/PMMA stacks were thoroughly rinsed in deionised water before being transferred to the substrate of choice. The samples were then dried in ambient conditions to ensure strong adhesion of the graphene film and, finally, the PMMA was removed in acetone.

It should be mentioned that there was a noticeable difference in the Raman spectra of samples prepared using different copper etchants. When ammonium persulfate was used, the spectra were characteristic of pristine graphene, with a narrow 2D peak (FWHM ~29 cm$^{-1}$). The samples had a very clean appearance, free of any contaminant particles. However, graphene transferred using this etchant sometimes had poor adhesion to the substrate and this method was not always suitable for device fabrication. On the other hand, samples transferred with iron chloride etching had a nearly 100% yield, however, in some cases the surface appeared less clean and the Raman spectra of single-layer graphene typically had a 2D peak with a FWHM of ~35 cm$^{-1}$, which is slightly larger than expected[S2].

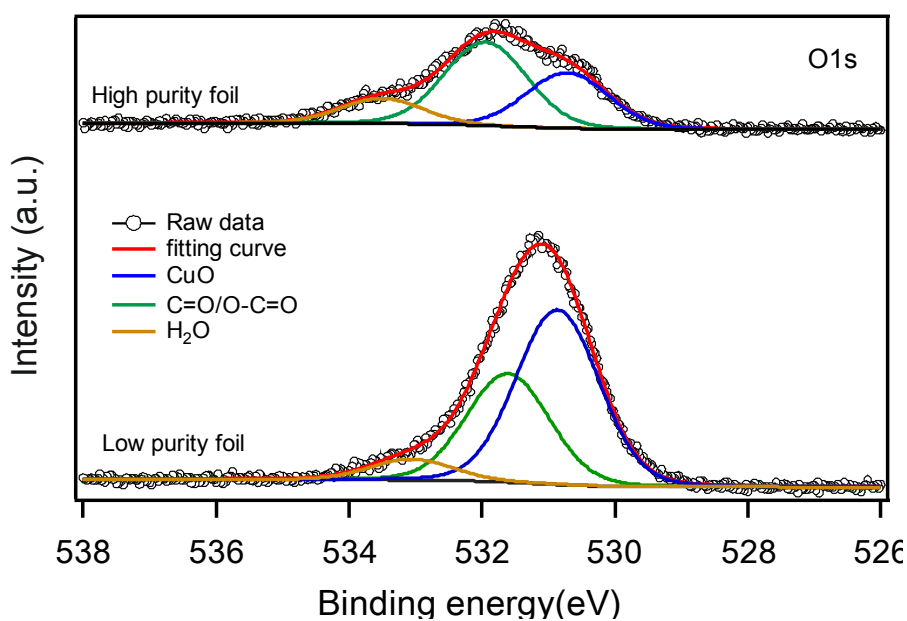

**Figure S6.** O1s core level spectra measured for high purity (top spectrum) and low purity (bottom spectrum) copper foil.

## 5. Additional XPS analysis

Detailed analysis of the O1s core level spectra measured for the higher purity (Sigma-Aldrich, 99.98%) and lower purity (Alfa Aesar, 99.8%) copper foils further confirms what was already discussed in the main text. While a similar amount of aerial contamination was found on the two foils, a significantly higher component attributed to cupric oxide was observed in the lower purity foil.

## 6. Estimation of the residual oxygen species in the reactor

The growth process is started only after thorough purging of the gas lines and of the reactor with Ar 6.0. At least two cycles of purging and pumping down to the base pressure of $10^{-3}$ mbar are done at the beginning of each process to eliminate atmospheric contamination as much as possible. The gases used have all a purity level of 6.0 (i.e., 99.9999%) and we can evince from the gas specifications that the total of oxygen containing species is < 0.2 ppm. Hence, the maximum partial pressure of oxygen containing species at the beginning of the process (and so throughout the entire growth) is (assuming worst case approximations) maximum $2.1 \times 10^{-13}$ atm. Now, making a plausible approximation for ideal gases we get that the concentration of oxygen species in the reactor is:

$$\frac{n}{V} = \frac{P}{RT} = \frac{0.21 \cdot 10^{-12}}{0.0821 \cdot 298} = 8.58 \cdot 10^{-15} \frac{\text{mol}}{\text{L}}$$

Therefore, in the total reactor volume of 21 L we approximately estimate the amount of oxygen species to be equal to $1.8 \times 10^{-13}$ mol, which translates to about $10^{11}$ oxygen molecules present during the process in the reactor. Considering the confined volume of the enclosure to be 0.005 L, the presence of oxygen molecules in the confined volume is about $2.6 \times 10^{7}$ molecules. To estimate the amount of oxygen introduced in the chamber by the oxidised foil, we calculate an estimated number of oxygen molecules for a typical sample size of 2x2 cm. The density of oxygen atoms in a Cu(111) crystal terminated with a layer of $Cu_2O$ (worst case assumption, as it presents a lower atomic density than CuO) is $6.4 \times 10^{14}$ atoms per cm2 (Ref. S3). Assuming that we have a single passivating layer, our foil – when introduced in the chamber – presents about $2.56 \times 10^{15}$ molecules, several orders of magnitude more than the potential oxygen residue in the reactor. Similar results can be obtained for $Cu_2O$ on Cu(100)[S3]. These rough estimates also support our assumption: it is the oxygen initially contained in the foil, rather than the oxygen residue in the reactor, which plays a relevant role in the formation of large grain graphene.

## 7. TEM mapping

During the initial development stages of the graphene growth process, TEM diffraction mapping described before[S4] was employed to determine the grain size of the small-grain polycrystalline films. In this method a parallel electron beam is scanned over an area of a few µm² while the electron diffraction patterns are collected by the Nanomegas ASTAR system with an external fast CCD. The result of the data collection is a set of several thousand patterns, each of which corresponds to a specific point in the sample. Using dedicated software each pattern is compared with a bank of simulated patterns of the crystal phases present in the sample and its crystallographic orientation is determined. As a final result, an orientation map in which the different grains can be identified is produced, with a resolution which depends on the size of the electron beam employed. In the case of graphene we had to use a large parallel beam of 150 nm to have a detectable diffraction signal of the fast CCD. The pattern matching for graphene is much simpler since we have to determine only the in-plane rotation of the graphite [0001] hexagonal pattern.

Electron-diffraction mapping provided us a quick analysis of the crystal size and orientation for small-grain continuous films. An example of an orientation map is displayed in Figure S7 (b). The colour gives the orientation of the hexagonal pattern with respect to the vertical direction of the image: blue correspond to (10-10) vertical while green correspond to (10-10) horizontal, light blue is intermediate between the two. A similar kind of mapping has been obtained using dark field images by Huang *et al*[S5].

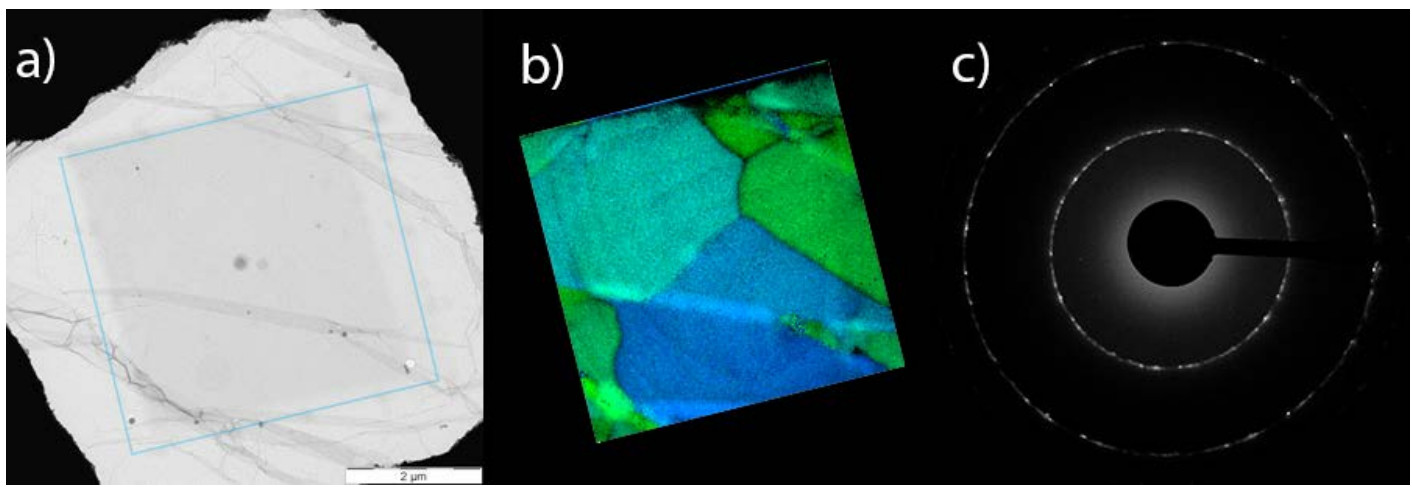

**Figure S7.** TEM mapping of polycrystalline graphene film. a) Bright field image of polycrystalline graphene film. b) Reconstructed map of the crystal orientations of the corresponding film. c) SAED pattern of a polycrystalline film. taken on an area of 4.5 µm in diameter.

While this diffraction mapping is limited to areas of a few microns square and is not suitable for the large-grain graphene, this technique collects diffraction patterns for each spot of the analysed area, making the data more comprehensive than the images obtained by dark-field mapping. This can be useful for mapping samples with very low crystal size and can distinguish grains with small rotational misalignment and therefore it could be applied in the development of other kinds of emerging 2D materials.

## Supplementary references: